\documentclass[epj]{svjour}

\usepackage{graphicx}
\usepackage{fancyhdr}
\usepackage{multirow}
\def\makeheadbox{{
 }}
\def\mytitle{My title} 
\def\myauthors{My name}  
\def\mytype{My type of session}
\def\mysession{My session}

\def\mytitle{Contact Interactions at the LHC} 
\def\myauthors{M\'onica V\'azquez Acosta}    
\def\mytype{Contributed Talk}    
\def\mysession{Alternatives}

\begin{document}
\title{Contact Interactions at the LHC}
\author{M\'onica L. V\'azquez Acosta (on behalf of the ATLAS and CMS Collaborations)\inst{}
}                     
%
%
\institute{CERN, G\`eneve 23, CH-1211 Geneva, Switzerland
}
%
\date{}
\abstract{
Contact interactions offer a general framework for describing a new  interaction
with a scale above the energy scale probed. These interactions can occur if the Standard Model  particles are
composite or if new heavy particles are exchanged. The discovery potential of contact
interactions at the LHC in dimuon and dijet final states at startup and  the asymptotic reach are presented.
} 
\maketitle
\section{Introduction}
\label{intro}
Quark compositeness or new interactions mediated by a new massive particle can be
approximated by a contact interaction, when the center-of-mass energy of the partons initiating the interaction, $\sqrt{s}$, is below the scale $\Lambda$~\cite{bib:subs,bib:subs2}. This is analogous to the effective four-fermion interaction which can describe the weak force at low energies.

\vspace*{-0.1cm}
\section{Dijet final state}
\label{sec:1} 
Quark compositeness in dijet events has been studied assuming a CI Lagrangian formed by the product of left-handed quark currents:

${\cal L} = \frac{2\pi A}{\Lambda^2}\sum_{i,j=1}^6(\bar{q_i}\gamma^\mu q_{iL})(\bar{q_j}\gamma^\mu q_{jL})$,
where $A=\pm 1$ which can give constructive or destructive interference with the Standard Model (SM).

Contact interactions will produce an increase in the event rate relative to QCD at high mass.
Observation of CI in mass distributions requires a precise understanding of QCD dijet cross sections,
 due to the large uncertainties in the jet energy scale and in the parton distribution functions (PDF)
at high mass. CI are expected to be more isotropic than the QCD background, since QCD is
dominated by the t-channel scattering and produces jets predominantly in the forward region.
Angular distributions have much smaller systematic uncertainties than cross sections measurements versus dijet mass.

\vspace*{-0.1cm}
\subsection{ATLAS Contact interaction sensitivity}
\label{sub:1}

\begin{figure}
\begin{center}
\includegraphics[width=0.48\textwidth,angle=0]{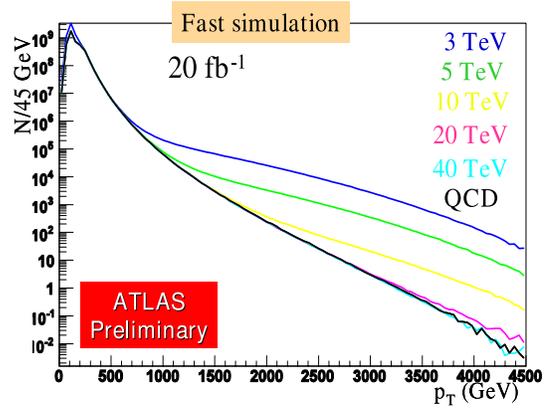}
\vspace*{-0.5cm}
\caption{$p_t$ distribution of the two leading jets showing the QCD prediction and the effect
of different quark compositeness scales}
\label{fig:a1}       
\end{center}
\end{figure}

\begin{figure}
\begin{center}
\includegraphics[width=0.48\textwidth,angle=0]{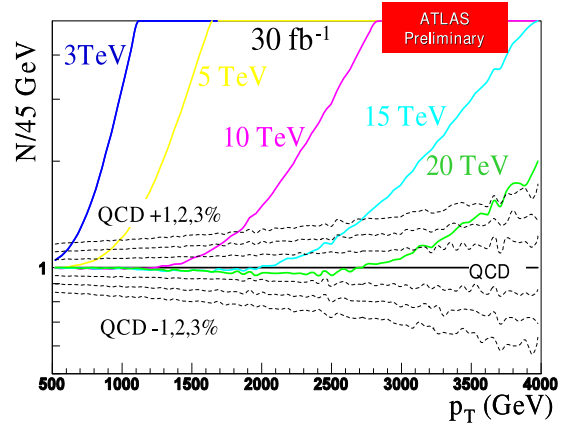}
\vspace*{-0.5cm}
\caption{Ratio of the $p_t$ distribution of the two leading jets for different compositeness scales to the QCD prediction}
\label{fig:a3}       
\end{center}
\end{figure}

\begin{figure}
\begin{center}
\includegraphics[width=0.45\textwidth,angle=0]{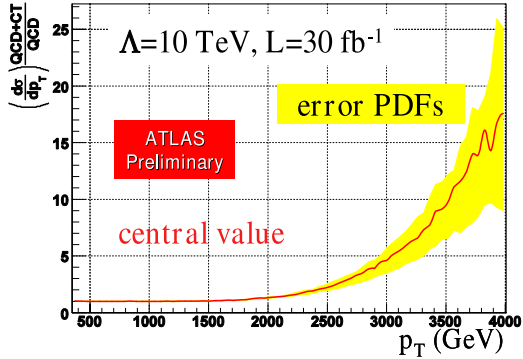}
\vspace*{-0.5cm}
\caption{PDF uncertainty in the ratio of the $p_t$ distribution of the two leading jets for a contact interaction scale of $\Lambda=10$ TeV to the QCD prediction}
\label{fig:a4}       
\end{center}
\end{figure}

\begin{figure}
 \begin{center}
\includegraphics[width=0.48\textwidth,angle=0]{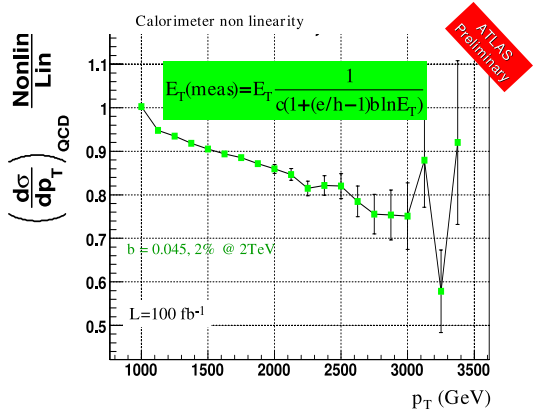}
\vspace*{-0.5cm}
\caption{Effect of calorimeter non linearities in the QCD dijet cross section}
\label{fig:a5}       
\end{center} \end{figure}

ATLAS has studied the effect of compositeness in the PTDR~\cite{bib:atlastdr}. New preliminary results with the ATLAS fast simulation have been produced, which include more recent PDFs. The $p_t$ distribution of the two leading jets has been studied (Fig.~\ref{fig:a1}). 
An uncertainty of $1\%$ is enough to hide CI scales of $\Lambda=20$~TeV (Fig.~\ref{fig:a3}). The uncertainties on 
the PDFs (Fig.~\ref{fig:a4}) and calorimeter non linearity (Fig.~\ref{fig:a5}) are large in the dijet cross section distributions.

\begin{figure} \begin{center}
\includegraphics[width=0.45\textwidth,angle=0]{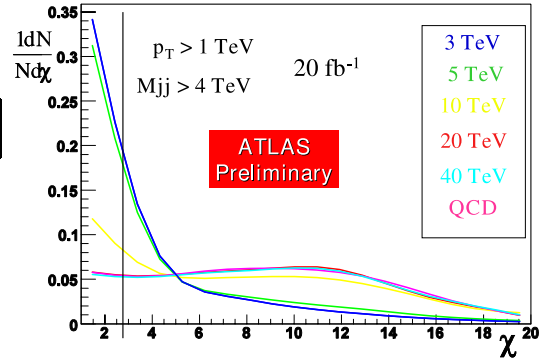}
\vspace*{-0.5cm}
\caption{Dijet Angular distribution  showing the QCD prediction and the effect
of different quark compositeness scales}
\label{fig:a6}       
\end{center} \end{figure}

The effect of CI in the dijet angular distribution versus $\chi=e^{|\eta_1-\eta2|}$, where $\eta_{1,2}$ are the pseudorapidities of the two leading jets has been studied (Fig.~\ref{fig:a6}).
For the $2\rightarrow 2$ parton scattering, it is related to the centre-of-mass scattering angle $\theta^*$ as follows: $\chi =\frac{1+|cos\theta^*|}{1-|cos\theta^*|}$. If one defines $R_\chi$ as the fraction of events with $R_\chi=\frac{N(\chi<\chi_{\rm cut})}{N(\chi>\chi_{\rm cut})}$ and the sensitivity as $R_1=\frac{R_{\chi}(\Lambda)-R_{\chi}(SM)}{\sqrt{\sigma^2_{\Lambda}+\sigma^2_{SM}}}$, the luminosity required to achieve a sensitivity of $R_1=3$ is presented in Table~\ref{tab:1}. The value of $\chi_{\rm cut}=2.8$ maximizes the sensitivity. Systematic uncertainties are expected to be much smaller than in the $d\sigma/dp_t$ case.

\begin{table}
\caption{ATLAS Preliminary: Luminosity to achieve a contact interaction sensitivity of~$R_1=3$ in dijet angular distributions. Systematic uncertainties are not included.}
\label{tab:1}       
\begin{tabular}{|c|c|c|c|c|c|}
\hline
$\Lambda$(TeV) & 3 & 5 & 10 & 20 & 40   \\
\hline
Lumi & $<1$ pb$^{-1}$ &  $6$ pb$^{-1}$ &  $0.7$ fb$^{-1}$ & $34$ fb$^{-1}$ & $426$ fb$^{-1}$ \\
\hline
\end{tabular}
\vspace*{1cm}  
\end{table}

\subsection{CMS Contact interaction sensitivity}
\label{sub:2}
\begin{figure} \begin{center}
\includegraphics[width=0.45\textwidth,angle=0]{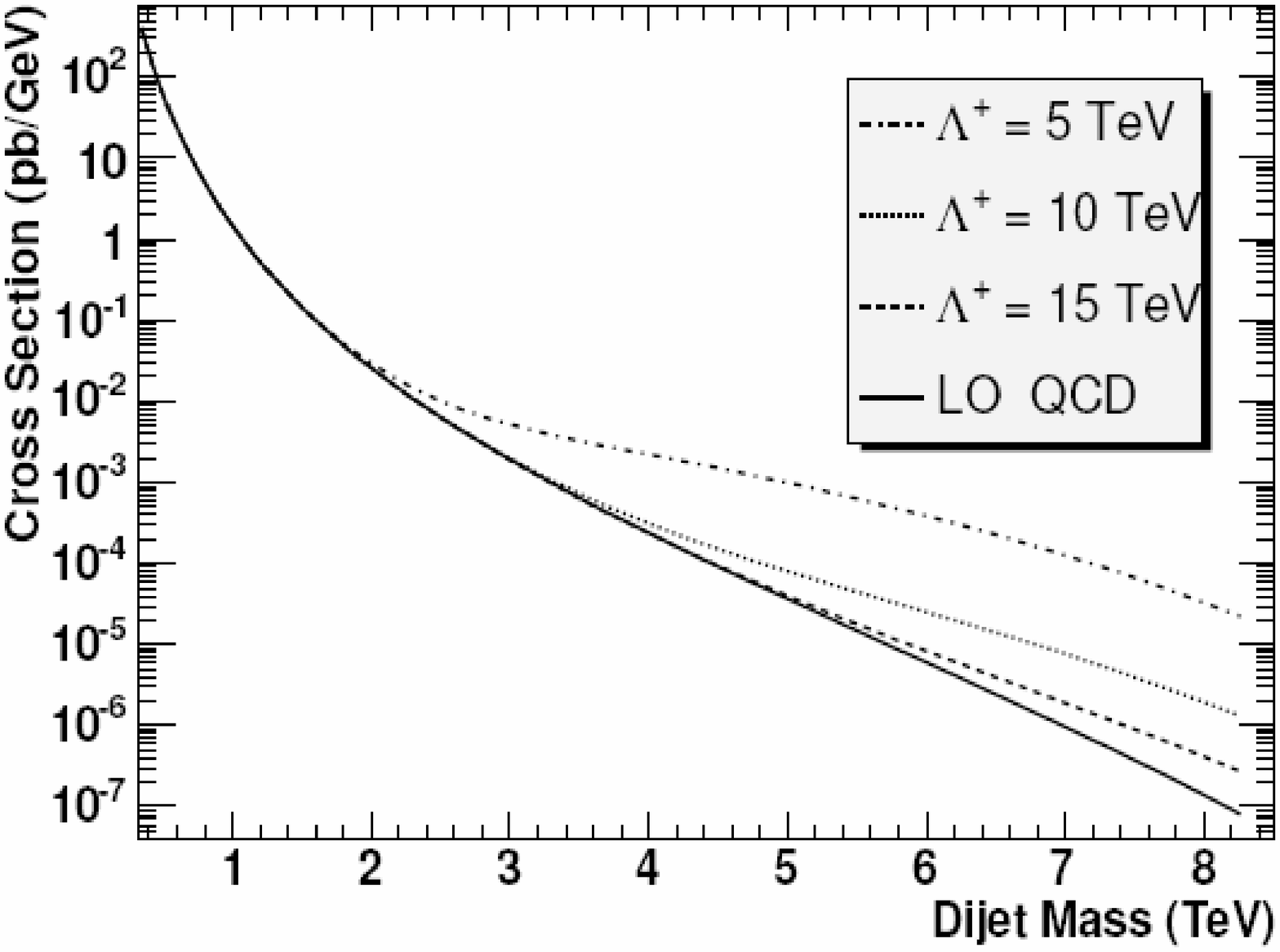}
\vspace*{-0.5cm}
\caption{CMS: Dijet mass cross section of the two leading jets showing the QCD prediction and the effect
of different contact interaction scales}
\label{fig:c1}       
\end{center} \end{figure}

\begin{figure} \begin{center}
\hspace*{-1.5cm}
\includegraphics[width=0.45\textwidth,angle=0]{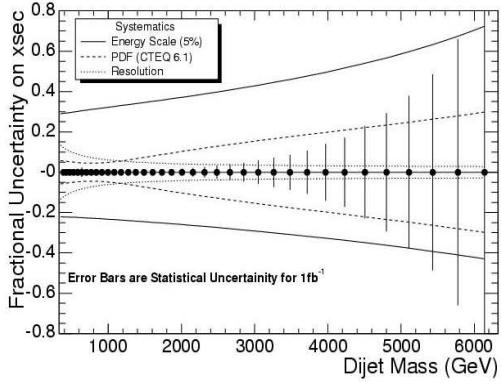}
\vspace*{-0.5cm}
\caption{CMS: Systematic uncertainties on the Dijet Mass cross section}
\label{fig:c2}       
\end{center} \end{figure}

CMS has studied the effects of compositeness~\cite{bib:cmstdr,bib:cmsjets}
 in the dijet cross section as a function of the dijet mass (Fig.~\ref{fig:c1}).
It is expected that a jet energy scale uncertainty of $\pm 5\%$ is achievable, which can produce changes in the dijet mass
cross section of $30-70\%$ (Fig.~\ref{fig:c2}).

\begin{figure} \begin{center}
\includegraphics[width=0.45\textwidth,angle=0]{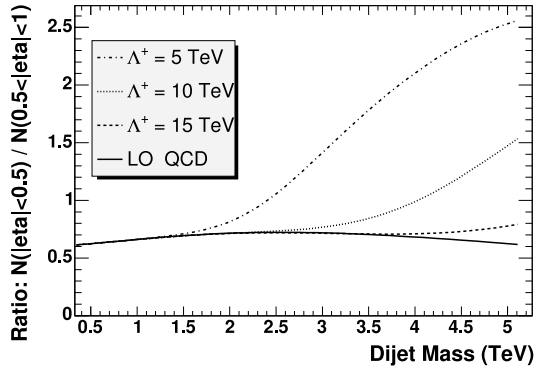}
\caption{CMS: Dijet ratio showing the QCD prediction and the effect
of different contact interaction scales as a function of the dijet mass}
\label{fig:c3}       
\end{center} \end{figure}

\begin{figure} \begin{center}
\includegraphics[width=0.45\textwidth,angle=0]{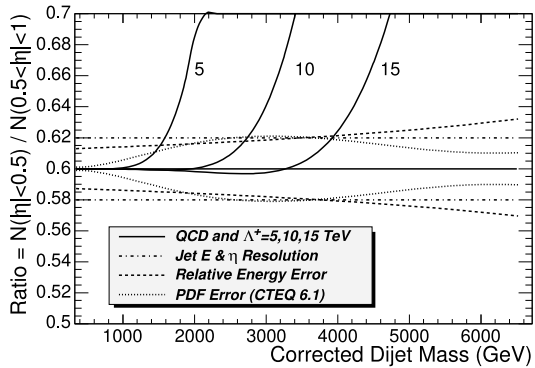}
\caption{CMS: Systematic uncertainties on the Dijet Ratio}
\label{fig:c4}       
\end{center} \end{figure}

The ratio of the number of dijets in which both jets have $|\eta|<0.5$ to the number of dijets in which both jets have $0.5<|\eta|<1$ as function of the dijet mass, is a simple measure of the most sensitive part of the angular distribution. The effects of CI in the dijet ratio have been studied (Fig.~\ref{fig:c3}) and the systematic uncertainties (Fig.~\ref{fig:c4}) have been to be much smaller than in the case of the dijet mass cross section. The CI scales that can be excluded at $95\%$ confidence level or can be discovered with a significance of $5\sigma$ are shown in Table~\ref{tab:2} for a luminosity of 100 pb$^{-1}$, 1 fb$^{-1}$ and 10 fb$^{-1}$. Scales up to 6.2 TeV can be excluded with a luminosity of 100 pb$^{-1}$. The D0 experiment has excluded scales up to 2.7 TeV~\cite{bib:d0} with an analysis that uses the same dijet ratio and a luminosity of 100 pb$^{-1}$.

\begin{table}
\caption{CMS: Contact interaction $95\%$ CL exclusion limits and $5\sigma$ discovery reach in dijet events, with the inclusion of statistical uncertainties only and
with all systematic uncertainties taken into account}
\label{tab:2}       
\begin{tabular}{|c|c|c|c|c|}
\hline
\multicolumn{2}{|c|}{Luminosity}  & 100 pb$^{-1}$ & 1 fb$^{-1}$ &  10 fb$^{-1}$  \\
\multicolumn{2}{|c|}{}  & $\Lambda$(TeV) & $\Lambda$(TeV) &  $\Lambda$(TeV)  \\
\hline
$95\%$ CL & Stat Only & 6.4 & 10.6 & 15.1 \\
Exclusion & All Syst & 6.2 & 10.4 & 14.8 \\
\hline
$5\sigma$& Stat Only & 4.7 & 8.0 & 12.2 \\
Discovery & All Syst & 4.7 & 7.8 & 12.0 \\
\hline
\end{tabular}
\vspace*{1cm}  
\end{table}


\section{Dimuon final state}
\label{sec:2}
Contact interaction in the dimuon final state have been studied assuming a non-parity conserving LL model. Contact interactions are expected to produce deviations from the Drell-Yan spectrum  at high dimuon invariant mass. 
\subsection{CMS Contact interaction sensitivity}
\label{sub:3}
\begin{figure} \begin{center}
\includegraphics[width=0.48\textwidth,angle=0]{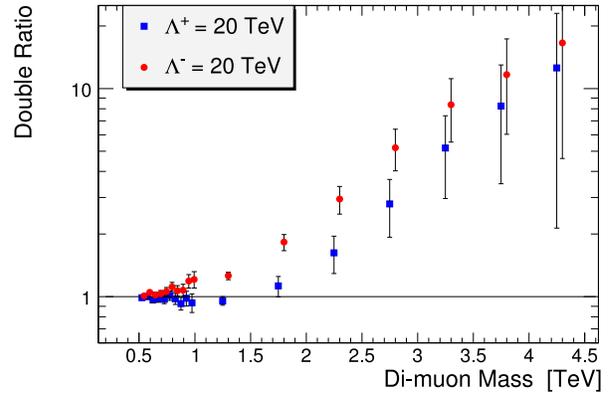}
\caption{CMS: Double ratio in the dimuon channel for contact interactions with a scale of $\Lambda=20$ TeV}
\label{fig:c5}       
\end{center} \end{figure}

CMS has studied the sensitivity to contact interaction in the dimuon final state~\cite{bib:cmstdr,bib:cmsmuons}.
A double ratio method has been developed to reduce systematic uncertainties. The ratio of the number of observed events in the dimuon mass bin i and a zeroth normalization bin  $R_i^{\rm data}=\frac{N_i^D}{N_0^D}=\frac{\sigma_i^D\epsilon_i^D}{\sigma_0^D\epsilon_0^D}$ is defined, where $\sigma$ is the cross section and $\epsilon$ is the experimental efficiency. The normalization bin is chosen to be between 250-500 GeV, above the $Z$ pole and in a region well covered by the Tevatron where the standard model has been seen to be valid. In this region the $u$ quark PDF is dominant which has the smallest uncertainties. A similar ratio is defined for the Monte Carlo simulation $R_i^{MC}=\frac{N_i^MC}{N_0^MC}=\frac{\sigma_i^{MC}\epsilon_i^{MC}}{\sigma_0^{MC}\epsilon_0^{MC}}$. The double ratio $DR_i=\frac{R_i^{\rm data}}{R_i^{MC}}$ is studied versus dijet mass and is shown for a scale of $\Lambda=20$ TeV in Fig.~\ref{fig:c5}. In the case of perfect theory understanding and detector modelling, a value of  $DR_i=1$ is expected. 

\begin{figure} \begin{center}
\includegraphics[width=0.45\textwidth,angle=0]{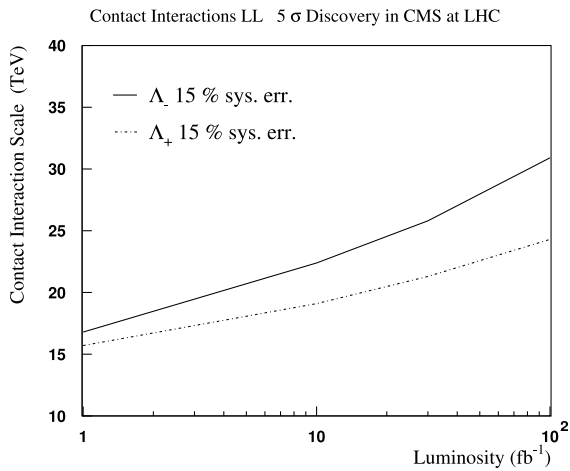}
\caption{CMS: $5\sigma$ discovery reach of contact interactions in the dimuon channel versus luminosity}
\label{fig:c6}       
\end{center} \end{figure}

\begin{figure} \begin{center}
\includegraphics[width=0.45\textwidth,angle=0]{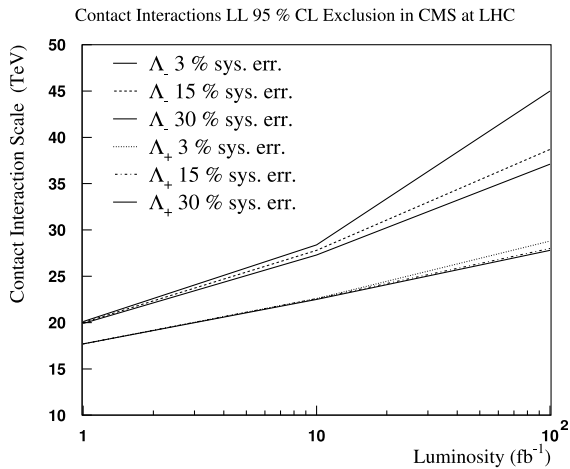}
\caption{CMS: $95\%$ CL exclusion limit of contact interactions in the dimuon channel versus luminosity}
\label{fig:c7}       
\end{center} \end{figure}

The $5\sigma$ discovery reach (Fig.~\ref{fig:c6}) and the $95\%$ CL exclusion limit (Fig.~\ref{fig:c7}) for contact interactions in the dimuon channel have been studied as a function of the luminosity. Up to a luminosity of 10 fb$^{-1}$ the measurements are dominated by statistical uncertainties. Systematic uncertainties of up to $30\%$ have a small impact in the discovery potential.

\section{Conclusions}
Contact interaction at a scale $\Lambda$ can be observed before any new exchanged particle
is directly seen. Many techniques have been developed to study compositeness and show promising results with low
systematic effects. The sensitivity of the ATLAS and CMS experiments to contact interactions
has been investigated. The first hundred pb$^{-1}$ of data will allow the discovery of contact
interactions with a scale up to $\Lambda = 5$ TeV. A luminosity of 100~fb~$^{-1}$ will allow
the discovery of compositeness up to $\sim$~30~TeV.

\end{document}